\documentclass[useAMS,usenatbib,usegraphicx]{mn2e}

\newcommand{\etalb}{et al.}
\newcommand{\beq}{\begin{equation}}
\newcommand{\beqa}{\begin{eqnarray}}
\newcommand{\eeq}{\end{equation}}
\newcommand{\eeqa}{\end{eqnarray}}

\newcommand{\br}{{\bf r}}
\newcommand{\bk}{{\bf k}}

\newcommand{\Lya}{Ly$\alpha$~}

\newcommand{\td}{{\tilde{\delta}}}
\newcommand{\tr}{^{\rm tr}}
\newcommand{\ap}{\alpha_\perp}
\newcommand{\arec}{a_{\rm rec}}

\title[Separating out the Alcock-Paczy\'{n}ski Effect on 21cm Fluctuations]
{Separating out the Alcock-Paczy\'{n}ski Effect on 21cm Fluctuations}

\author[R. Barkana]{R. Barkana
\thanks{E-mail: barkana@wise.tau.ac.il}\\
School of Physics and Astronomy, The Raymond and Beverly Sackler
Faculty of Exact Sciences,\\ Tel Aviv University, Tel Aviv 69978,
ISRAEL}

\begin{document}

\pagerange{\pageref{firstpage}--\pageref{lastpage}} \pubyear{2005}

\maketitle

\label{firstpage}

\begin{abstract}
We reconsider the Alcock-Paczy\'{n}ski effect on 21cm fluctuations
from high redshift, focusing on the 21cm power spectrum. We show that
at each accessible redshift both the angular diameter distance and the
Hubble constant can be determined from the power spectrum, at epochs
and on scales where the ionized fraction fluctuations are
linear. Furthermore, this is possible using anisotropies that depend
only on linear density perturbations and not on astrophysical sources
of 21cm fluctuations. We show that measuring these quantities at high
redshift would not just confirm results from the cosmic microwave
background; if the 21cm power spectrum can be measured to better than
$10\%$ precision, it will improve constraints from the CMB alone on
cosmological parameters including dark energy.
\end{abstract}


\section{Introduction}

Resonant absorption by neutral hydrogen at its spin-flip 21cm
transition can be used to map its three-dimensional distribution at
early cosmic times \citep{Hogan, Scott, Madau}. The primordial
inhomogeneities in the cosmic gas induced variations in the optical
depth for absorption of the cosmic microwave background (CMB) at the
redshifted 21cm wavelength. Absorption occurs as long as the spin
temperature of hydrogen, $T_s$ (which characterizes the population
ratio of the upper and lower states of the 21cm transition), is lower
than the CMB temperature, $T_\gamma$.  This condition is satisfied in
the redshift interval $20\la z\la 200$ \citep{Loeb04}, before the
first galaxies formed in the universe \citep{BL01}, while 21cm
emission is expected at later epochs.

Several groups are currently constructing low-frequency radio arrays
capable of detecting the diffuse 21cm radiation; these include the
{\it Primeval Structure Telescope} (web.phys.cmu.edu/$\sim$past), the
{\it Mileura Widefield Array} (web.haystack.mit.edu/arrays/MWA), and
the {\it Low Frequency Array} (www.lofar.org). The 21cm measurements
depend on the spin temperature $T_s$ which itself depends on the
kinetic gas temperature $T_k$ and also on the strength of coupling
between $T_s$ and $T_k$, a coupling that is due to atomic collisions
or to indirect coupling through Ly$\alpha$ radiation from the first
stars \citep{Wout, Field, Madau, Ciardi}. Fluctuations in these
quantities are expected to produce isotropic 21cm
fluctuations. Upcoming experiments are expected to successfully detect
the 21cm fluctuations despite the required high sensitivity and the
presence of strong foregrounds \citep{Zalda04, Santos, Miguel}.

Cosmological measurements generally determine three-dimensional
locations indirectly, using angular position on the sky and redshift
along the line of sight. This indirect route can modify any measured
power spectrum and produce spurious anisotropies even when the
underlying power spectrum is intrinsically isotropic. One such effect
is that of redshift distortions, which appear when peculiar velocity
gradients cause apparent changes in line-of-sight distances and thus
in apparent densities. This effect has been studied extensively in
galaxy redshift surveys \citep{kaiser, nd94, th96, des04}, where it
can be used to probe a degenerate combination of the cosmic mean
density of matter and galaxy bias. In the case of fluctuations in the
21cm brightness temperature relative to the CMB, the same effect
\citep{Indian, Indian2} produces a line-of-sight anisotropy which is
of great importance since it permits a separation of the physics from
the astrophysics of 21cm fluctuations \citep{BLlos}; i.e., it allows
observers to measure fundamental cosmological parameters through the
linear power spectrum, and to separately probe the effects of stellar
and quasar radiation on the intergalactic medium.

A second anisotropy results if data are analyzed using assumed
cosmological parameters that differ from the true ones. In this
situation, separations along the line of sight are scaled differently
from those on the sky, and this distorts the appearance of any
spherical object or isotropic statistical distribution. This creates a
spurious anisotropy that the Alcock-Paczy\'{n}ski (AP) test exploits
in order to constrain the cosmological parameters \citep{AP, Hui}.

Recently, \citet{Nusser} suggested to apply the AP test to 21cm
fluctuations from high redshift. \citet{Nusser} considered 21cm
fluctuations produced by density perturbations and neutral fraction
fluctuations, as well as the anisotropies due to redshift distortions
and the AP effect. \citet{Nusser} showed that the correlation function
(as a function of comoving position $\br$) has a portion $P_6(\mu_r)$
that appears only in the presence of AP anisotropy, where $\mu_r$ is
the cosine of the angle between $\br$ and the line of sight, and $P_6$
is the 6$^{\rm th}$-order Legender polynomial. \citet{APindian} also
analyzed anisotropies in the clustering pattern of redshifted 21cm
maps. They considered the same anisotropy sources as \citet{Nusser}
and showed that the anisotropy is affected both by the H I
distribution and by cosmological parameters, making it hard to
disentangle the AP effect; however, they did not include the
$P_6(\mu_r)$ term but only considered lower-order terms.

In this paper we reconsider the AP effect on 21cm fluctuations from
high redshift. Our analysis is valid even in the presence of the whole
gamut of possible sources of 21cm fluctuations: isotropic fluctuations
in gas density, \Lya flux \citep{21gal}, neutral fraction, and
temperature, as well as the velocity gradient anisotropy and the AP
effect. We do not assume that these sources produce fluctuations that
are proportional to density fluctuations, we do not neglect
temperature fluctuations, and we also do not assume the Einstein-de
Sitter growing mode when calculating the redshift distortions;
instead, we apply exact linear perturbation theory at high redshifts,
including the effects of the CMB and the difference in the evolution
of the dark matter and the baryons \citep{BLbinf, Smadar}. At lower
redshifts, when the first galaxies produce \Lya, UV, and X-ray
photons, we assume only that the 21cm fluctuations remain small and
can be treated to linear order. Fluctuations that are due to optically
thin radiation backgrounds are expected to indeed be small, since any
given gas parcel encounters radiation that is averaged over sources
distributed in a large surrounding region, leading to linear
fluctuations as in the case of \Lya flux \citep{21gal}. Thus our
results are valid at high redshift before the formation of the first
galaxies, and later during \Lya coupling and X-ray heating.  Such a
regime of observable 21cm fluctuations before reionization is indeed
expected theoretically \citep{Madau, Sethi, Fur}.

During reionization, our results are valid if X-rays reionized the
universe \citep{oh01, mr04}. Stellar UV photons instead produce
sharply-defined H II regions, but in this case the neutral fraction
fluctuations may still be small on large enough scales, especially
early in reionization, when the cosmic mean neutral fraction is still
much smaller than unity. There are several difficulties, however. Once
reionization begins in earnest, comoving scales well above 10 Mpc may
be required in order to exceed the typical size of ionized bubbles
\citep{BLflucts, FurHII1}. On such large scales, the correlations are
relatively weak and thus harder to observe, and also the delay due to
the light travel time along the line of sight produces a significant
additional anisotropy on such scales \citep{mcquinn, BLlight}.

Unlike previous studies of the AP effect we focus on the 21cm power
spectrum, a quantity which is much closer to the observable radio
visibilities. We first consider the AP effect on power spectra in
general (\S~2), and then show that the 21cm power spectrum allows for
a simple separation of the AP anisotropy from other effects, as well
as a separate measurement of the Hubble constant at high redshift
(\S~3). We show that the appropriate power spectrum coefficients are
large enough to be feasible to measure (\S~4.1), and the measured
quantities yield extra cosmological information beyond CMB
measurements (\S~4.2). We summarize our results in \S~5.

\section{The AP Effect on the Power Spectrum}

An incorrect choice of cosmological parameters in the analysis of
cosmological data scales both the angular and line-of-sight
coordinates. We denote the angular distance out to $z$ as $D_A$, and
the Hubble constant at $z$ as $H$. Following \citet{Nusser}, we define
$(1+\alpha)$ as the ratio of the true value of $H D_A$ to its assumed
value. This quantity depends on the ratio between the line-of-sight
and angular scales, and thus introduces anisotropy in the power
spectrum and corresponds to the classical AP test. However, even a
redshift-dependent overall scaling of the coordinates (i.e., with
$\alpha=0$) may have observable effects on the power spectrum, as
discussed below. Thus, we also define $(1+\ap)$ as the same ratio but
for $D_A$ rather than for the quantity $H D_A$. This implies that any
assumed angular distance equals $1/(1+\ap)$ times the true distance,
and any assumed line-of-sight distance equals $(1+\alpha)/ (1+\ap)$
times the true one. Overall, we refer to these various scalings simply
as the AP effect.

Suppose we observe the relative fluctuation $\delta$ in some
underlying field (which may be the density of some component or the
21cm brightness temperature $T_b$). Then if we observe $\delta$ at a
comoving position $\br$ (typically measured in Mpc) with coordinates
$(x_c,y_c,z_c)$, where $\br$ is reconstructed from the angular
position and redshift using the assumed cosmology, we actually detect
$\delta$ at a true comoving position $\br\tr$ with coordinates \beq
\br\tr = \left((1+\ap)x_c, (1+\ap)y_c, \frac{1+\ap} {1+\alpha} z_c
\right)\ .\eeq If the Fourier transform of $\delta(\br\tr)$ with
respect to $\br\tr$ (i.e., in the absence of the AP effect) is
$\td\tr(\bk)$, where the comoving wavevector $\bk$ has coordinates
$(k_x, k_y, k_z)$, then the observed Fourier transform of
$\delta(\br)$ with respect to $\br$ is \beq \td(\bk) = \frac{1+\alpha}
{(1+\ap)^3} \td\tr(\bk\tr)\ , \eeq where we have defined \beq \bk\tr
\equiv \left(\frac{1} {1+ \ap} k_x, \frac{1} {1+\ap} k_y,
\frac{1+\alpha} {1+\ap} k_z \right)\ .\eeq Using the definition of the
power spectrum in terms of an ensemble average and a Dirac delta
function, we define power spectra $P(\bk)$ and $P\tr(\bk)$ as \beq
\langle \td (\bk_1) \td (\bk_2) \rangle = (2\pi)^3
\delta^D(\bk_1+\bk_2) P(\bk_1)\ , \eeq and \beq \langle \td\tr (\bk_1)
\td\tr (\bk_2) \rangle = (2\pi)^3 \delta^D(\bk_1+\bk_2) P\tr(\bk_1)\ ,
\eeq respectively. We then find that the AP effect distorts the
observed power spectrum of $\delta$ compared with the true one: \beq
P(\bk) = \frac {1+\alpha} {(1+\ap)^3} P\tr(\bk\tr)\ .  \eeq

We now specialize to an underlying power spectrum of the form
$P\tr(k\tr, \mu\tr)$ where $\mu\tr = k_z\tr/k\tr$ is the cosine of the
angle between the $\bk$ vector and the line of sight; thus, we assume
a power spectrum with a line-of-sight anisotropy only. We also expand
to first order in the corrections $\alpha$ and $\ap$. Then using the
relations \beq \frac{k\tr}{k} = 1+\alpha \mu^2 - \ap\ ;\ \
\frac{\mu\tr}{\mu} = 1+ \alpha(1- \mu^2)\ , \eeq we obtain \beqa
P(k,\mu) & = & (1+\alpha -3 \ap) P\tr + (\alpha \mu^2 - \ap)\,
\frac{\partial P\tr} {\partial \log k} \label{eq:AP} \\ & & +\,
\alpha(1- \mu^2)\, \frac{\partial P\tr} {\partial \log \mu}\ ,
\nonumber \eeqa where henceforth $P\tr$ with no explicit argument is
evaluated at the {\em observed}\/ $\bk$.

\section{The case of the 21cm power spectrum}

We assume a flat universe with cosmological parameters
\citep{CMB} $h=0.72$, $\Omega_m=0.27$, and $\Omega_r=8.0 \times 10^{-5}$
(assuming 3 massless neutrinos), with the remaining energy density in
a cosmological constant. Then the mean brightness temperature offset
on the sky at redshift $z$ is
\citep{Madau} \beqa
T_b & = & 4.74\, {\rm mK}\, \left[ \frac{(1+z)^2} {H(z)/H_0}
\right]\, \left( \frac{\Omega_b h} {.033}\right) \left({{T_s - 
T_{\gamma}} \over {T_s}}\right) \\ & \simeq & 28.8\, {\rm mK}\,
\left( \frac{\Omega_b h} {.033} \right) \left( \frac{\Omega_m} {.27} 
\right)^{- \frac{1} {2}} \left({{1+z}\over{10}}\right)^{1 \over 2} 
\left({{T_s - T_{\gamma}}\over {T_s}}\right)\ , \nonumber 
\eeqa
where the second expression is accurate to $1\%$ at $4 \la z \la 70$
(where the contributions of both vacuum energy and radiation can be
neglected).

Since fluctuations in the 21cm brightness temperature are produced by
various isotropic sources plus the line-of-sight velocity gradients,
the linear power spectrum (in the absence of the AP effect) can be
written in the form \citep{BLlos} \beq P\tr_{T_b}(\bk) = \mu^4
P\tr_{\mu^4}(k) + \mu^2 P\tr_{\mu^2}(k) + P\tr_{\mu^0}(k) \ . \eeq
While $P\tr_{\mu^0}$ and $P\tr_{\mu^2}$ depend not only on
fluctuations in the gas density and temperature but also on stellar
radiation fields, $P\tr_{\mu^4}(k)$ is due to velocity gradients and
is always the power spectrum of $\dot{\delta}_b H^{-1}$, where
$\dot{\delta}_b$ is the time derivative of the baryon density.

To calculate the AP effect in this case we apply Eq.~(\ref{eq:AP})
and find a modified 21cm power spectrum:
\beq P_{T_b}(\bk) = \mu^6 P_{\mu^6}(k) + \mu^4 P_{\mu^4}(k) +
\mu^2 P_{\mu^2}(k) + P_{\mu^0}(k) \ , \eeq where \beqa 
P_{\mu^6} & = & - \alpha \left( 4 P\tr_{\mu^4} - \frac{\partial
P\tr_{\mu^4}} {\partial \log k} \right)\ ; \nonumber \\ P_{\mu^4} & =
& P\tr_{\mu^4} + \alpha \left( 5 P\tr_{\mu^4} -2 P\tr_{\mu^2} +
\frac{\partial P\tr_{\mu^2}} {\partial \log k}
\right) \nonumber \\ & & - \ap \left( 3 P\tr_{\mu^4} + \frac{\partial 
P\tr_{\mu^4}} {\partial \log k} \right)\ ; \label{eq:P4} \\
P_{\mu^2} & = & P\tr_{\mu^2} + \alpha \left(3 P\tr_{\mu^2}
+ \frac{\partial P\tr_{\mu^0}} {\partial \log k} \right)
-\ap \left(3 P\tr_{\mu^2} + \frac{\partial P\tr_{\mu^2}} 
{\partial \log k} \right)\ ; \nonumber \\
P_{\mu^0} & = & (1+\alpha) P\tr_{\mu^0} -  \ap \left(3 
P\tr_{\mu^0} +\frac{\partial P\tr_{\mu^0}} {\partial \log k}
\right)\ . \nonumber \eeqa

First of all, the AP effect changes the three observable isotropic
power spectra, $P_{\mu^4}$, $P_{\mu^2}$, and $P_{\mu^0}$. The effect
of an overall scaling (i.e., non-zero $\ap$) is to distort the shape
and normalization of each of these power spectra. On the other hand,
the parameter $\alpha$ corresponds to an anisotropy, and it mixes the
different power spectra. In particular, $P\tr_{\mu^4}(k)$ now gets a
correction of order $\alpha$ that can depend on stellar
radiation. However, the velocity distortion and AP anisotropies also
combine to produce a fourth observable power spectrum, $P_{\mu^6}$
[analogous to the $P_6$ term in the correlation function
\citep{Nusser}]. The $k$-dependence of this term depends on linear
perturbation theory only, and the term is not zero since the power
spectrum of density fluctuations has a logarithmic slope well below 4
on all scales. Thus, the new $P_{\mu^6}$ term is a direct probe of the
AP anisotropy parameter $\alpha$. Its coefficient, though, depends on
the linear power spectrum which is itself a function of the unknown
cosmological parameters. 

If we assume that the shape of the primordial power spectrum from
inflation can be described with a few parameters (e.g., with a power
law index plus its first derivative with respect to $k$), then these
parameters can be reconstructed at the same time as the cosmological
parameters in an iterative procedure. After the first measurement of
$\alpha$ is made at a given redshift, the cosmological parameters can
be adjusted to give the right value of $H D_A$ at that redshift. A
re-analysis of the data with the new cosmological parameters should
yield a new $\alpha$ that is close to zero, but not exactly, since
changing the parameters changes the $k$-dependent coefficient of
$\alpha$. The procedure can be iterated until $\alpha$ converges very
close to zero. At this point, $P_{\mu^4}$ allows a measurement of
$\ap$ from the predicted shape of $P\tr_{\mu^4}$. Again, the
measurement can be iterated until cosmological parameters are found
such that the 21cm power spectrum is consistent with both $\alpha$
and $\ap$ equal to zero.

This procedure is made easier by observing the 21cm power spectrum
(and thus the coefficients of $\alpha$ and $\ap$) at a range of values
of $k$ and at a range of redshifts, since the $k$ and $z$ dependence
of these terms is precisely known for given cosmological parameters.
Since the number of parameters in the standard cosmological models is
relatively small, a large number of measured points should help
overcome degeneracies as well as difficulties with noise and
foregrounds. The determination of the parameters may be helped by the
presence of distinct features in $P\tr_{\mu^4}$, specifically the
signatures of the acoustic oscillations on large scales (see
\S~4.2). Note that any change of $\ap$ with redshift further helps to
separate it from uncertainties in the shape of the power spectrum,
since the redshift dependence of the power spectrum itself is known
theoretically. The redshift variation of $\ap$ is however fairly weak,
e.g., if a $10\%$ overestimate of $\Omega_m$ is used in the analysis,
then $\ap=3.3\%$ at redshift 7, $3.5\%$ at $z=10$, $3.6\%$ at $z=20$,
and $3.8\%$ at $z=100$.

\section{Implications}

\subsection{Power spectra}

As noted in the previous section, the AP effect allows $\alpha$ and
$\ap$ to be measured from observations of 21cm fluctuations, and once
they are measured at a given redshift, a reanalysis with parameters
that set $\alpha=\ap=0$ can be used to recover the three true
underlying power spectra. The analysis is particularly simple at high
redshift, before the first galaxies formed (but at $z < 150$), on
large-scale structure scales (i.e., $k$ between 0.01 and 40
Mpc$^{-1}$). \citet{BLbinf} showed that in this regime the 21cm power
spectra at various redshifts can be described as linear combinations
of five fixed initial fluctuation modes (i.e., functions of $k$), with
redshift-dependent coefficients. Neglecting the AP effect, they argued
that the five modes and the linear coefficients can all be determined
directly from observations. Specifically, if the three observable
power spectra are measured at $N$ $k$-values at $M$ different
redshifts, then there are a total of $3NM$ data points. Modeling the
data in terms of the amplitudes of the five modes at the same $N$
$k$-values, along with two coefficients per redshift for each mode
[see \citet{BLbinf} for details], yields a total of $5N+10M$
parameters. In this case, $M=2$ redshifts suffice to determine all
the parameters as long as $N \ge 20$.

Now we consider adding to this the AP effect, within the theoretical
framework presented in the previous section but without assuming a
theoretical calculation of the modes or the redshift coefficients. We
thus add the two AP parameters per redshift, and the derivatives with
respect to $\log k$ of the five modes [see Eq.~(\ref{eq:P4})], for a
total of $10N+12M$ parameters. This must be compared with $4NM$ data
points, where we have included the $\mu^6$ term at each $k$ and
$z$. In this case, obtaining at least as many data points as
parameters requires $M=4$ redshifts if $N \ge 8$, while $M=3$
redshifts suffice if $N \ge 18$. Measurements at additional redshifts
or at more values of $k$ will allow a more accurate reconstruction
along with multiple consistency checks.

\begin{figure}
\includegraphics[width=84mm]{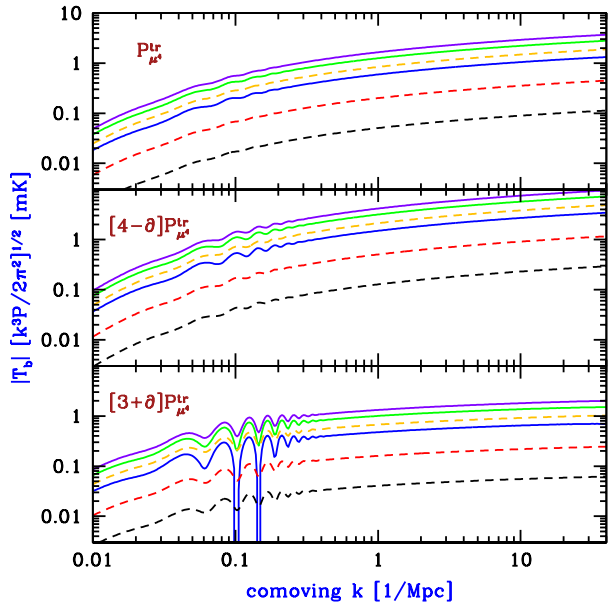}
\caption{Power spectrum coefficients of 21cm brightness fluctuations 
versus wavenumber. We show the quantity $|T_b| [k^3 P/(2
\pi^2)]^{1/2}$, where $P$ is replaced by $P\tr_{\mu^4}$ (upper panel),
$[4 - \partial_{\log k}]P\tr_{\mu^4}$ (middle panel), or $[3 +
\partial_{\log k}]P\tr_{\mu^4}$ (lower panel). In each case we consider
redshifts 150, 100, 50 (solid curves, from bottom to top), 35, 25, and
20 (dashed curves, from top to bottom), assuming no stellar radiation
is present.}
\label{fig:coeffs}
\end{figure}

\begin{figure}
\includegraphics[width=84mm]{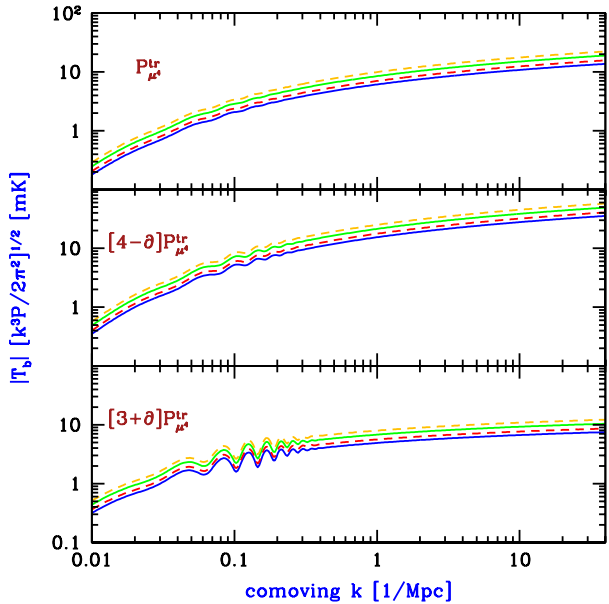}
\caption{Power spectrum coefficients of 21cm brightness fluctuations 
versus wavenumber. Same as Figure~\ref{fig:coeffs}, except that we
assume that $T_s \gg T_{\gamma}$, and we consider in each case
redshifts 20, 15, 10, and 7 (from bottom to top).}
\label{fig:coeffs2}
\end{figure}

Of more general interest are the two quantities involved in the
determination of $\alpha$ and $\ap$ at any redshift, even when
reionization and other complex processes are underway. The first is
the coefficient of $\alpha$ in $P_{\mu^6}$ in Eq.~(\ref{eq:P4}), which
can be used to determine $\alpha$. The second is the coefficient of
$\ap$ in $P_{\mu^4}$ in Eq.~(\ref{eq:P4}), which can be used to
determine $\ap$ after $\alpha$ has been zeroed.
Figure~\ref{fig:coeffs} shows $P\tr_{\mu^4}$ and these two
coefficients at high redshifts (in brightness temperature units),
assuming that galaxies have not yet formed and using the concordance
cosmological parameters. Note that the signature of the large-scale
acoustic oscillations in the power spectrum $P\tr_{\mu^4}$ is
magnified in its derivative $\partial_{\log k}P\tr_{\mu^4}$ and
therefore in the two coefficients. At some lower redshift, stellar
radiation is expected to heat the gas well above the CMB through
X-rays and to fully couple the spin temperature to the gas temperature
indirectly through \Lya photons; although there are large
uncertainties, projections of the properties of high-redshift galaxies
suggest that heating and spin coupling are likely to occur well before
cosmic reionization \citep{Madau, Sethi,
Fur}. Figure~\ref{fig:coeffs2} shows $P\tr_{\mu^4}$ and the same two
coefficients at lower redshifts assuming that $T_s \gg T_{\gamma}$ due
to stellar radiation. When considering the magnitudes of these terms,
it is important to note the spherical averages $\langle \mu^4
\rangle=1/5$ and $\langle \mu^6 \rangle=1/7$, although the sensitivity
of upcoming radio arrays may not be isotropic.

\subsection{Cosmological parameters}

CMB measurements \citep{CMB} have provided a precise determination of
features in the linear power spectrum, including the matter-radiation
turnover and the signature of the acoustic oscillations of the
baryon-photon fluid. These features measure the properties of the
various cosmic fluids at recombination, when dark energy (or a
cosmological constant) was negligible. However, the observed angular
scale subtended by these features provides a precise measurement of
the angular diameter distance $D_A$ to the epoch of recombination, an
integrated quantity that is affected by dark energy at low redshift.
The signatures of the acoustic oscillations have also been detected
recently in the large-scale distribution of massive galaxies at $z\sim
0.35$ \citep{eisen, surveys2}. Upcoming larger redshift surveys out to
$z \sim 1$ will begin to constrain significantly the properties of
dark energy \citep{hh03}, although the precise constraint depends on
prior assumptions on the possible complexity of the properties of dark
energy \citep{ram02}.

We now consider the role that 21cm measurements can play in the
general campaign to nail down the cosmological parameters and the
properties of dark energy. As shown above, the AP effect on the 21cm
power spectrum allows a direct measurement of $\alpha$ and $\ap$, or
equivalently of $H$ and of $D_A$, at applicable redshifts. In this
section we generalize to a flat universe with the energy density
$\Omega_w$ (which remains after accounting for matter and radiation)
consisting of a dark energy component with a constant equation of
state $p=w \rho$. We also assume cosmic recombination (decoupling) at
$\arec=1/1090$ \citep{CMB}, where $a=1/(1+z)$ is the scale
factor. Then the Hubble constant at scale factor $a$ is \beq H(a) =
H_0 \left[ \Omega_m a^{-3} + \Omega_r a^{-4} +\Omega_w a^{-3(1+w)}
\right]^{\frac{1}{2}}\ , \eeq and the angular diameter distance from
redshift zero is \beq D_A(a) = a \int_a^1 \frac{da'} {a'^2 H(a')}\
. \label{eq:DA} \eeq Although $D_A(a)$ at high redshift is sensitive
to the value of $w$, this sensitivity is for the most part due to the
low-redshift portion of the integral in Eq.~(\ref{eq:DA}). Thus, the
sensitivity is substantially the same as that of $D_A(\arec)$ which is
measured from the CMB. So in order to accurately assess the ability of
21cm measurements to contribute cosmological information beyond the
CMB, we consider the following fractional quantities: \beq f_H(a)
\equiv \frac{H(a)} {H(\arec)}\ ;\ \ f_D(a) \equiv \frac{D_A(\arec) -
\frac{\arec}{a} D_A(a)} {D_A(\arec)}\ . \eeq We consider derivatives
of $\log f_H$ (i.e., fractional changes in $f_H$), since $f_H \ll 1$
simply means that $H(a) \ll H(\arec)$ and this does not represent a
loss of precision. On the other hand, we consider derivatives of
$f_D$, since $D_A(a) \simeq D_A(\arec)$, and $f_D \ll 1$ represents a
real loss of precision in the subtraction in the numerator of $f_D$.

\begin{figure}
\includegraphics[width=84mm]{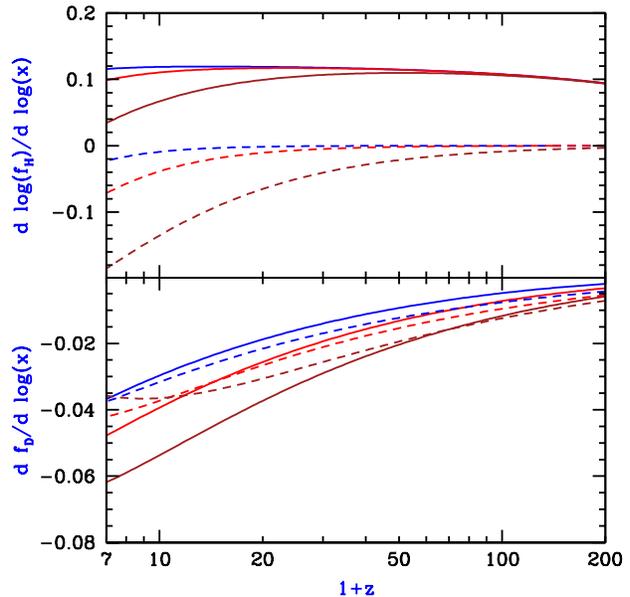}
\caption{Relative sensitivity of $\log f_H(a)$ and $f_D(a)$ to
cosmological parameters. For variations in a parameter $x$, we show
$d/d{\rm log}(x)$ of $\log f_H$ (top panel) and of $f_D$ (bottom
panel). In each case we consider variations in $\Omega_m$ (solid
curves) and in $w$ (dashed curves). When we vary each parameter we fix
the other one, and the variations are all carried out with respect to
the $\Omega$ values given in the text, in a flat $\Omega_{\rm
total}=1$ universe, for $w=-1$, -0.75, and -0.5 (from top to bottom in
each set of curves at $z=20$).}
\label{fig:cosmo}
\end{figure}

Figure~\ref{fig:cosmo} shows the dependence of $\log f_H$ and of $f_D$
on cosmological parameters, through their derivatives with respect to
$\log \Omega_m$ (in a flat universe with fixed $\Omega_r$), and with
respect to $\log w$, both as a function of $1+z$, for various values
of $w$ with the $\Omega$ values as given above. The relative
sensitivity to $\Omega_m$ is around $10\%$ for $\log f_H$ and 4--6$\%$
(at the low-redshift end) for $f_D$. The relative sensitivity to $w$
is significant only at the low redshift end ($z \sim 6$), where it is
$\sim 4\%$ for $f_D$ and between $2\%$ (if $w=-1$) and $20\%$ (if
$w=-0.5$) for $\log f_H$.

\section{Summary}

We have analyzed the AP effect on the power spectrum of 21cm
fluctuations at high redshift. Including various sources of 21cm
fluctuations, and assuming only that the fluctuations in the ionized
fraction are linear, we have shown that the AP effect produces an
anisotropic $\mu^6$ term that can be used to measure and eliminate the
AP anisotropy parameter $\alpha$. This leaves a contribution to the
$\mu^4$ term that can be used to measure and eliminate the second AP
parameter, $\ap$. Both these AP terms depend only on linear density
perturbations and not on 21cm fluctuations due to astrophysical
sources that are far more uncertain. Measuring the angular diameter
distance and the Hubble constant at high redshift (using $\alpha$ and
$\ap$) would provide constraints on $\Omega_m$ and $w$. Assuming the
same cosmological quantities are also measured at cosmic recombination
from the CMB, the 21cm measurements maintain a relative sensitivity of
order $10\%$ to the cosmological parameters even after the common
dependence with the CMB is eliminated. The planned 21cm experiments
(see \S~1) are expected to reach a brightness temperature sensitivity
$\la 1$ mK at $z \sim 10$, assuming that the foregrounds can be
removed effectively; observations at higher redshifts will be much
harder since the foreground emission increases approximately as
$(1+z)^{2.5}$.

\section*{Acknowledgments}
The author acknowledges support by NSF grant AST-0204514 and Israel
Science Foundation grant 28/02/01.

\bsp

\label{lastpage}

\end{document}